# Rare Decays of *K* and *B* Mesons Revisited

Swee Ping Chia
spchia@um.edu.my

*Physics Department, University of Malaya*
*50603 Kuala Lumpur, Malaysia*

The rare decays of mesons are higher order effects in weak interactions. The main contributions to such decay processes are from the *Z*-penguin, the photon-penguin and the box diagram. At the quark level, these contributions are well known. However, when considering the corresponding physical hadronic processes, a knowledge of how hadronic states are expressed in terms of quarks. In this paper, we suggest a simplistic way to relate the quark level process to the hadronic process. For simplicity, we consider here only the decays of pseudoscalar mesons. The meson-quark coupling is described in terms of a $\gamma_5$ coupling with a constant coupling constant. We apply the model to processes in which meson $M_1$ decays to meson $M_2$ with the emissions of neutrino pair and charged lepton pair. In calculating the decay rates, a cut-off momentum $\Lambda$ is introduced to control the divergent triangle integral over the internal momentum. The results of our calculation are compared to experimental data on the decay rates for *K*- and *B*-meson decays.

**Keywords:** Rare decays, electroweak process, *K* meson, *B* meson.

**PACS:** 12.15.Ji, 14.40.Aq, 14.40.Nd

## I. INTRODUCTION

One of the factors contributing to the success of the Standard Model is the presence of the weak neutral current mediated by $Z^o$ boson. This weak neutral current gives rise to a neutral change of quark flavours at higher orders in the weak interactions. Of special interest is the processes of the types $q_1 \to q_2 \nu \bar{\nu}$ and $q_1 \to q_2 \ell^+ \ell^-$. Such processes at the quark level are well-understood [1-18]. They are dominated by the penguin diagrams (both *Z*-penguin and photon-penguin) and the box diagram.

The corresponding physical processes which involve hadrons, namely $M_1 \to M_2 \nu \bar{\nu}$ and $M_1 \to M_2 \ell^+ \ell^-$, are more involved. This is because a good understanding of how hadronic states are described in terms of quarks is lacking.

We suggest here a simplistic way to relate the quark level process to the hadron process. For simplicity, we consider here only the decays of pseudoscalar mesons. The meson-quark coupling is described in terms of a $\gamma_5$ coupling as follow in Fig. 1. The Hadronic processes are then depicted as shown in Fig. 2.

## II. THE QUARK LEVEL VERTEX FUNCTIONS

The calculation at the quark level is straight-forward. Within the Standard Model, the process with neutrino pair production $q_1 \to q_2 \nu \bar{\nu}$ receives contributions from the Z-penguin and the box diagram. For the process with charged lepton pair production $q_1 \to q_2 \ell^+ \ell^-$, the photon-penguin also contributes in addition to the Z-penguin and the box diagram. In the following, we present both the processes in a unified way.

The combined contribution from the penguins and the box diagram at the lepton vertex is:

$$\Gamma_\mu = \frac{-g^2}{32\pi^2 M_W^2} \gamma_\mu [L S^L(x_j) + R S^R(x_j)], \tag{1}$$

where $j$ is the internal quark line, $x_j = m_j^2 / M_W^2$, and $S^L(x_j)$ and $S^R(x_j)$ are the dynamical form factors from the penguins and the box diagram [19-22]. For $q_1 \to q_2 \nu \bar{\nu}$,

$$S^L(x_j) = -\frac{1}{2} g^2 f(x_j), \tag{2}$$

$$S^R(x_j) = 0, \tag{3}$$

$$f(x) = f^{(B)}(x) + f^{(Z)}(x), \tag{4}$$

$$f^{(B)}(x) = 4x(x-1-\ln x)/(x-1)^2, \tag{5}$$

$$f^{(Z)}(x) = \frac{1}{2} x[(-6+x)(x-1)+(2+3x)\ln x]/(x-1)^2. \tag{6}$$

For $q_1 \to q_2 \ell^+ \ell^-$,

$$S^{L,R}(x_j) = \frac{1}{8} g^2 \tilde{S}^{L,R}(x_j), \tag{7}$$

$$\tilde{S}^L(x_j) = f^{(B)}(x_j) - 16\sin^2\theta_W f^{(\gamma)}(x_j) - 4(2\sin^2\theta_W - 1)f^{(Z)}(x_j), \tag{8}$$

$$\tilde{S}^R(x_j) = -16\sin^2\theta_W f^{(\gamma)}(x_j) - 8\sin^2\theta_W f^{(Z)}(x_j), \tag{9}$$

$$f^{(\gamma)}(x) = \frac{1}{72}[x^2(25-19x)(x-1) \\ - 2(8-32x+54x^2-30x^3+3x^4)\ln x]/(x-1)^4. \tag{10}$$

The dynamical form factors are functions of $x_j$ only, where $j$ is the internal quark flavour in the penguins and the box diagram. For our calculation, we use the following values for the $u$-, $c$-, and $t$-quarks [23]:

$$m_u = 0.0023 \text{ GeV}, \quad m_c = 1.275 \text{ GeV}, \quad m_t = 173.5 \text{ GeV}. \tag{11}$$

With this set of values for the internal quark masses, the dynamical form factors for the decays with neutrino pair production can be calculated from Eqs. (2) to (6), and those for decays with charged lepton pair production can be calculated from Eqs. (7) to (10). The results of the calculation are shown in Table 1.

TABLE 1. Dynamical form factors.

| Internal quark $j$ | $x_j$ | $f(x_j)$ | $\tilde{S}^L(x_j)$ | $\tilde{S}^R(x_j)$ |
| --- | --- | --- | --- | --- |
| u | 8.1866×10⁻¹⁰ | 5.0569×10⁻⁸ | −17.197 | −17.197 |
| c | 2.5158×10⁻⁴ | 6.0057×10⁻³ | −6.8073 | −6.8093 |
| t | 4.6583 | 6.3749 | 11.241 | −5.4054+00 |

## III.  THE DECAY RATE

The calculation of the decay amplitude $M$ is straightforward:

$$M = (-1)\int \frac{d^4k}{(2\pi)^4} \left(\frac{-g^2}{32\pi^2 M_W^2}\right) \sum_j \lambda_j \bar{v}\gamma^\mu [LS^L(x_j) + RS^R(x_j)]u \times$$

$$\times Tr\left[\gamma_\mu L \frac{i(\gamma.k - m_1)}{k^2 - m_1^2} ig_{M1}\gamma_5 \frac{i(\gamma.(P_1 - k) - m)}{(P_1 - k)^2 - m^2} ig_{M2}\gamma_5 \frac{i(\gamma.(P_1 - P_2 - k) - m_2)}{(P_1 - P_2 - k)^2 - m_2^2}\right],$$

(12)

where $g_{M1}$ and $g_{M2}$ are the coupling constants at the $M_1$-quark and $M_2$-quark vertices. The integration over the internal momentum $k$ is logarithmically divergent. A cut off momentum $\Lambda$ is introduced. Assuming that this cut-off momentum is large compared to all the masses involved, we get:

$$M = \left(\frac{-g^2}{32\pi^2 M_W^2}\right) g_{M1} g_{M2} \left(\frac{i}{32\pi^2} \ln\frac{\Lambda^2}{M^2}\right)\left(-\frac{10}{3}\right) \sum_j \lambda_j \bar{v}\gamma.P_1[LS^L(x_j) + RS^R(x_j)]u \,. \quad (13)$$

The decay rate is then obtained by integrating over the phase space of $\langle |M|^2 \rangle$. After some algebra, we obtain the following expressions: For $M_1 \to M_2 \nu\bar{\nu}$,

$$\Gamma = \left(\frac{25}{9}\frac{G_F^2 M_W^4}{8(32)^4 \pi^{11}}\right)\left(g_{M1} g_{M2} \ln\frac{\Lambda^2}{M_W^2}\right)^2 M_1^5 R(y) \left|\sum_j \lambda_j f(x_j)\right|^2, \quad (14)$$

where $\lambda_j = V_{2j}V_{1j}^*$, $y = M_2^2/M_1^2$, and $R(y)$ is the kinematic factor given by

$$R(y) = (1-y)(1 - 7y - 7y^2 + y^3) - 12y^2 \ln y \,. \quad (15)$$

And for $M_1 \to M_2 \ell^+ \ell^-$,

$$\Gamma = \left(\frac{25}{27}\frac{G_F^2 M_W^4}{4(32)^5 \pi^{11}}\right)\left(g_{M1} g_{M2} \ln\frac{\Lambda^2}{M_W^2}\right)^2 M_1^5 R(y)\left\{\left|\sum_j \lambda_j \tilde{S}^L(x_j)\right|^2 + \left|\sum_j \lambda_j \tilde{S}^R(x_j)\right|^2\right\}. (16)$$

We are interested in the following decays: $K \to \pi$, $B \to \pi$, and $B \to K$. The kinematic factors as calculated from Eq. (15) are tabulated in Table 2:

Table 2. Kinematic factor $R(y)$ for each of the three decay transitions.

| Transition | $y$ | $R(y)$ |
|---|---|---|
| $K \to \pi$ | 7.9928×10⁻² | 0.55832 |
| $B \to \pi$ | 6.9894×10⁻⁴ | 0.99445 |
| $B \to K$ | 8.7446×10⁻³ | 0.93440 |

## IV.  CKM MATRIX ELEMENTS

The CKM matrix elements play an important role in our analysis. We take from the 2012 Review of Particle Physics the following magnitudes [24]:

$$V_{ud} = 0.97427(15), \quad V_{us} = 0.22534(65), \quad V_{ub} = 0.00351(15),$$
$$V_{cd} = 0.22520(65), \quad V_{cs} = 0.97344(16), \quad V_{cb} = 0.0412(8),$$
$$V_{td} = 0.00867(30), \quad V_{ts} = 0.0404(8), \quad V_{tb} = 0.999146(34). \quad (17)$$

What enter our calculation are the products of CKM matrix elements $\lambda_j = V_{2j} V_{1j}^*$. The phases of the CKM elements are not well determined. However, we take advantage of the unitarity triangle relation offered by

$$\lambda_u + \lambda_c + \lambda_t = 0 \quad (18)$$

to have a good estimate of the relative phases among the three $\lambda_j$.

For $s \to d$ transition:
$$\lambda_u = 0.21954$$
$$\lambda_c = -0.21919 + i\, 0.13570 \times 10^{-3}$$
$$\lambda_t = -0.32336 \times 10^{-3} - i\, 0.13570 \times 10^{-3} \quad (19)$$

For $b \to d$ transition:
$$\lambda_c = 0.92782 \times 10^{-2}$$
$$\lambda_t = -0.80528 \times 10^{-2} + i\, 0.31926 \times 10^{-2}$$
$$\lambda_u = -0.12254 \times 10^{-2} - i\, 0.31926 \times 10^{-2} \quad (20)$$

For $b \to s$ transition:
$$\lambda_t = 0.40365 \times 10^{-1}$$
$$\lambda_c = -0.40099 \times 10^{-1} + i\, 0.74462 \times 10^{-3}$$
$$\lambda_u = 0.27119 \times 10^{-3} - i\, 0.74462 \times 10^{-3} \quad (21)$$

## V. THE DYNAMICAL FORM FACTORS

As shown by Eqs. (14) and (15), the decay rates for the processes $M_1 \to M_2 \nu \bar{\nu}$ and $M_1 \to M_2 \ell^+ \ell^-$ respectively involve the following dynamical form factors $H_{(0)}$, $H^{(L)}$ and $H^{(R)}$ defined by:

$$H_{(0)} = \left| \sum_j \lambda_j f(x_j) \right|^2, \quad (22)$$

$$H^{(L)} = \left| \sum_j \lambda_j \tilde{S}^L(x_j) \right|^2, \quad (23)$$

$$H^{(R)} = \left| \sum_j \lambda_j \tilde{S}^L(x_j) \right|^2. \quad (24)$$

Taking advantage of the relative phases of the various $\lambda_j$ factors, estimates for $H_{(0)}$, $H^{(L)}$ and $H^{(R)}$ are obtained as shown in Table 3.

Table 3. Values for the dynamical form factors $H_{(0)}$, $H^{(L)}$ and $H^{(R)}$.

| Transition | $H_{(0)}$ | $H^{(L)}$ | $H^{(R)}$ |
|---|---|---|---|
| $s \to d$ | $1.2157 \times 10^{-6}$ | 5.2295 | 5.2391 |
| $b \to s$ | $6.6092 \times 10^{-2}$ | 0.53495 | $3.6019 \times 10^{-3}$ |
| $b \to d$ | $2.6523 \times 10^{-3}$ | $2.5827 \times 10^{-2}$ | $1.4192 \times 10^{-3}$ |

## VI. COMPARISON WITH EXPERIMENTAL DATA

Our calculation is applied to the following processes:

$K \to \pi e^+ e^-$
$K \to \pi \nu \bar{\nu}$
$B \to K e^+ e^-$
$B \to K \nu \bar{\nu}$
$B \to \pi e^+ e^-$
$B \to \pi \nu \bar{\nu}$

Comparison with available experimental decay rates is shown in Table 4. By comparing to Experimental decay rates, we can obtain an estimate for the parameter $g_{M1} g_{M2} \ln(\Lambda^2 / M_W^2)$.

Table 4. Comparison of the calculated decay rates with experimental rates.

| | |
|---|---|
| $K \to \pi e^+ e^-$ | $\Gamma^{(cal)} = [g_K g_\pi \ln(\Lambda^2 / M_W^2)]^2 (3.1057 \times 10^{-24} MeV)$ |
| | $\Gamma^{(exp)} = (1.595 \pm 0.048) \times 10^{-20} MeV$ |
| | $g_K g_\pi \ln(\Lambda^2 / M_W^2) = 71.7 \pm 1.1$ |
| $K \to \pi \nu \bar{\nu}$ | $\Gamma^{(cal)} = [g_K g_\pi \ln(\Lambda^2 / M_W^2)]^2 (1.7312 \times 10^{-28} MeV)$ |
| | $\Gamma^{(exp)} = (9.0 \pm 5.8) \times 10^{-24} MeV$ |
| | $g_K g_\pi \ln(\Lambda^2 / M_W^2) = 228 \pm 73$ |
| $B \to K e^+ e^-$ | $\Gamma^{(cal)} = [g_B g_K \ln(\Lambda^2 / M_W^2)]^2 (3.7393 \times 10^{-20} MeV)$ |
| | $\Gamma^{(exp)} = (2.21 \pm 0.28) \times 10^{-16} MeV$ |
| | $g_B g_K \ln(\Lambda^2 / M_W^2) = 76.9 \pm 4.9$ |
| $B \to K \nu \bar{\nu}$ | $\Gamma^{(cal)} = [g_B g_K \ln(\Lambda^2 / M_W^2)]^2 (2.2027 \times 10^{-19} MeV)$ |
| | $\Gamma^{(exp)} < 5.21 \times 10^{-16} MeV$ |
| | $g_B g_K \ln(\Lambda^2 / M_W^2) < 48.6$ |
| $B \to \pi e^+ e^-$ | $\Gamma^{(cal)} = [g_B g_\pi \ln(\Lambda^2 / M_W^2)]^2 (1.9931 \times 10^{-21} MeV)$ |
| | $\Gamma^{(exp)} < 3.21 \times 10^{-17} MeV$ |
| | $g_B g_K \ln(\Lambda^2 / M_W^2) < 127$ |
| $B \to \pi \nu \bar{\nu}$ | $\Gamma^{(cal)} = [g_B g_\pi \ln(\Lambda^2 / M_W^2)]^2 (9.4077 \times 10^{-21} MeV)$ |
| | $\Gamma^{(exp)} < 4.01 \times 10^{-14} MeV$ |
| | $g_B g_K \ln(\Lambda^2 / M_W^2) < 2065$ |

The results obtained can be said to be satisfactory. Only three decay modes, out of six, are experimentally measured, namely: $K \to \pi e^+ e^-$, $K \to \pi \nu \bar{\nu}$ and $B \to K e^+ e^-$ [23]. For the $K \to \pi$ transitions, the value of 71.7±1.1 obtained for $g_K g_\pi \ln(\Lambda^2/M_W^2)$ from $K \to \pi e^+ e^-$ is quite accurate. But for $K \to \pi \nu \bar{\nu}$, the value obtained is 228±73, which is a factor of 3 too high. There are two factors worth mentioning here. First, the experimental rate for $K \to \pi \nu \bar{\nu}$ has a large error. Second, our result rely heavily on the relative phases determined using the unitarity triangle. Such phases would involve huge error because some of the CKM mixing elements are not well-determined. The second factor is not accounted for in the error for $g_{M1} g_{M2} \ln(\Lambda^2/M_W^2)$.

For the $B \to K$ transitions, the value for $g_B g_K \ln(\Lambda^2/M_W^2)$ is 76.9±4.9 from $B \to K e^+ e^-$. But for $B \to K \nu \bar{\nu}$, the value obtained is <47.6, which is more than 5 standard deviations away from the value obtained from $B \to K e^+ e^-$.

## VII. CONCLUSION

What we have presented here is an attempt to extend calculation of electroweak processes at the quark-level to physical processes involving hadrons. The basic assumption is the $\gamma_5$ coupling at the meson-quark vertex. This assumption has been applied earlier to calculate $K^o \to$ vacuum amplitude [25]. The calculation involves assuming a cut-off momentum $\Lambda$ to overcome the logarithmic divergent in the integration over the quark loop. To relate our calculation to the experimental decay rates, we have to exploit the CKM mixing elements, which are quite well measured. Through utilizing the unitarity relation, we are able to obtain estimates for the relative phases for the products of CKM elements $\lambda_j$. The calculation gives reasonable agreement with experimental values. The values for the parameter $g_{M1} g_{M2} \ln(\Lambda^2/M_W^2)$ can be used in other similar processes, such as flavour-changing decays of Higgs boson, flavour-changing decays of Z boson, and production of pseudoscalar meson pair of different flavours from $e^+ e^-$ collision.

## ACKNOWLEDGMENTS

This research was carried out under the Research Grant UMRG049/09AFR from University of Malaya. Part of this research was carried out while on research leave at Academia Sinica, Taiwan. The author wishes to thank Prof. Hai-Yang Cheng for valuable discussion.

Fig. 1. The pseudoscalar meson coupling to quarks is approximated by a constant $\gamma_5$ coupling.

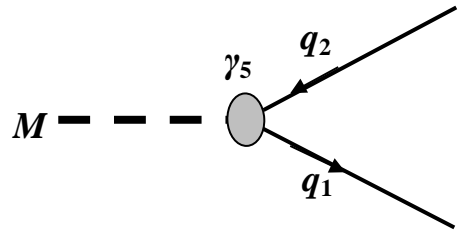

Fig. 2. Diagram depicting the decays $M_1 \to M_2 \nu \bar{\nu}$ and $M_1 \to M_2 \ell^+ \ell^-$ utilizing the constant $\gamma_5$ coupling at the meson-quark vertices. $\Gamma_\mu$ is the vertex function of the penguin and box diagrams.

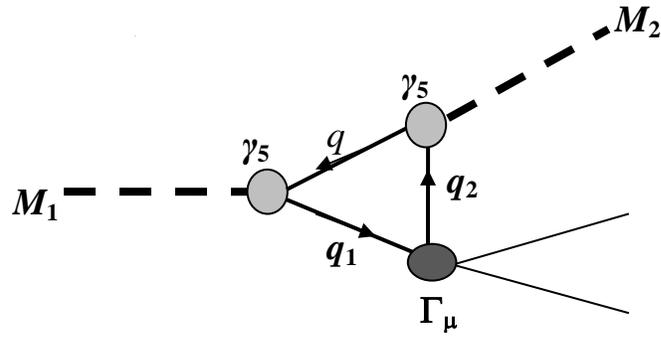